# Transparent Data Encryption- Solution for Security of Database Contents

Dr. Anwar Pasha Abdul Gafoor Deshmukh [1]
[1]College of Computers and Information Technology, University of Tabuk, Tabuk 71431, Kingdom of Saudi Arabia
[1]mdanwardmk@yahoo.com

Dr. Riyazuddin Qureshi [2]
[2]Community College, Jazan University, Jazan, Kingdom of Saudi Arabia
[2]riaz_quraysh21@yahoo.com

*Abstract*— The present study deals with Transparent Data Encryption which is a technology used to solve the problems of security of data. Transparent Data Encryption means encrypting databases on hard disk and on any backup media. Present day global business environment presents numerous security threats and compliance challenges. To protect against data thefts and frauds we require security solutions that are transparent by design. Transparent Data Encryption provides transparent, standards-based security that protects data on the network, on disk and on backup media. It is easy and effective protection of stored data by transparently encrypting data. Transparent Data Encryption can be used to provide high levels of security to columns, table and tablespace that is database files stored on hard drives or floppy disks or CD's, and other information that requires protection. It is the technology used by Microsoft SQL Server 2008 to encrypt database contents. The term encryption means the piece of information encoded in such a way that it can only be decoded read and understood by people for whom the information is intended. The study deals with ways to create Master Key, creation of certificate protected by the master key, creation of database master key and protection by the certificate and ways to set the database to use encryption in Microsoft SQL Server 2008.

## I. INTRODUCTION

Present day life is vastly driven by Information Technology. Extensive use of IT is playing vital role in decision-making in all commercial and non-commercial organizations. All activities are centered on data, its safe storage and manipulation. In present scenario organizations' survival is at stake if its data is misused. Data is vulnerable to a wide range of threats like, Weak Authentication, Backup Data Exposure, Denial of Service, etc.

This study is aimed at to deal with the most critical of those threats to which database is vulnerable. Transparent Data Encryption (TDE) shields database up to considerable extent against such threats. TDE is used to prevent unauthorized access to confidential database, reduce the cost of managing users and facilitate privacy managements. This latest technology arms users' i.e. database administrators to solve the possible threats to security of data. This technology allows encrypting databases on hard disk and on any backup media. TDE nowadays, is the best possible choice for bulk encryption to meet regulatory compliance or corporate data security standards.

Main purpose of Transparent Data Encryption is to provide security to columns, tables, Tablespace of database. In section 1, we explain about the Encryption, Plaintext, Cipher text (Encrypted Text) with simple examples and Types of Encryption. Section 2 explains Transparent Data Encryption, its Scope, Uses and its Limitations. Section 3 deals with Transparent Data Encryption in Microsoft Server 2008 with its architecture. Section 4 addresses issues like Create Database Using Microsoft SQL Server 2008 with coding and its description; Section 5 is for Creation of Encryption and Decryption of Database with coding and its descriptions respectively. Section 6 deals with conclusion.

## II. ENCRYPTION

Encryption is said to occur when data is passed through a series of mathematical operations that generate an alternate form of that data; the sequence of these operations is called an algorithm. To help distinguish between the two forms of data, the unencrypted data is referred to as the plaintext and the encrypted data as cipher text. Encryption is used to ensure that information is hidden from anyone for whom it is not intended, even those who can see the encrypted data. The process of reverting cipher text to its original plaintext is called decryption. This process is illustrated in the Figure 1 below.

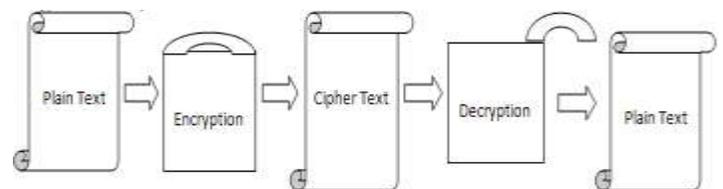

Figure 1 Process of Encryption and Decryption

A. Simple Example

- Encryption of the word "abcd" could result in "wxyz" Reversing the order of the letters in the plaintext generates the cipher text.
- It is Simple encryption and quit easy for attacker to retrieve the original data. A better solution of

encrypting this message is to create an alternative alphabet by shifting each letter by arbitrary number.
- Example: The word "abcd" is encrypted by shifting the alphabet by 3 letters to right then the result will be "defg". The "Ceasar Cipher" means exchange of letters or words with another.
- Normal Alphabet: abcdefghijklmnopqrstuvwxyz
- Alphabet shifted by 3: defghijklmnopqrstuvwxyzabc

*B. Types of Encryptions*

There are two general categories for key-based encryption
- Symmetric
- Asymmetric

*1) Symmetric*

Symmetric encryption uses a single key to encrypt and decrypt the message. To encrypt the message provide key to the recipient before decrypt it. To use symmetric encryption, the sender encrypts the message and, if the recipient does not already have a key, sends the key and cipher text separately to the recipient. The message then decrypt by recipient key. This method is easy and fast to implement but has weaknesses; for instance, if a hacker intercepts the key, he can also decrypt the messages. Single key encryptions tend to be easier for hacker /cracker. This means that the algorithm that is used to encode the message is easier for attackers to understand, enabling them to more easily decode the message.

*2) Asymmetric encryption*

Asymmetric encryption, also known as Public-Key encryption, uses two different keys - a public key to encrypt the message, and a private key to decrypt it. The public key is used to encrypt and private key to decrypt. One can easily distribute the public key to communicate because only with private key one can decrypt it. To protect the message between users the sender encrypts it by public key. Then receiver decrypts it by private key. Only recipient can decrypt the message in this type of encryptions.

III. TRANSPARENT DATA ENCRYPTION

Transparent data encryption is used for encryption and decryption of the data and log files. The encryption uses a Database Encryption Key (DEK), which is stored in the database boot record for availability during recovery. It is Asymmetric key secured by using a certificate stored in the master database. Transparent data encryption protects data and log files. It is a technology used to solve the problems of security of data means encrypting databases on hard disk and on any backup media. Transparent Data Encryption can be used to provide high levels of security to columns, table and Tablespace that is database files stored on hard drives or floppy disks or CD's, and other information that requires protection. It is the technology used by Microsoft SQL Server 2008 to encrypt database contents.

Transparent data encryption encrypts data before it's written to disk and decrypts data before it is returned to the application. The encryption and decryption process is performed at the SQL layer, completely transparent to applications and users. Subsequent backups of the database files to disk or tape will have the sensitive application data encrypted.

*A. Scope*

Extensive use of IT is playing vital role in decision-making in commercial as well as non-commercial organizations. In present scenario operations of any organization are badly affected if; data is misused. Data is vulnerable to a wide range of threats like, Excessive Privilege Abuse, Legitimate Privilege Abuse, Weak Authentication, Backup Data Exposure, etc.

The crucial role data plays in any organization itself explains its importance. So it is exposed to grave threats of theft, misuse or loss. This study is aimed to deal with the most critical of those threats to which database is vulnerable by focusing on Transparent Data Encryption (TDE). TDE is used to prevent unauthorized access to confidential database, reduce the cost of managing users, and facilitate privacy managements. This latest technology enables users' i.e. database administrators to counter the possible threats to security of data. TDE facilitates encrypting databases on hard disk and on any backup media. TDE nowadays, is the best possible choice for bulk encryption to meet regulatory compliance or corporate data security standards.

*B. Use of Transparent data encryption*

There are three important uses of Transparent data encryption as below

1) Authentication
2) Validation
3) Data Protection

*1) Authentication*

Unauthorized access to information is a very old problem. Business decisions today are driven by information gathered from mining terabytes of data. Protecting sensitive information is key to a business's ability to remain competitive. Access to key data repositories such as the Microsoft SQL Server 2008 that house valuable information can be granted once users are identified and authenticated accurately. Verifying user identity involves collecting more information than the usual user name and password. Microsoft SQL Server 2008 Advanced Security provides the ability for businesses to leverage their existing security infrastructures such as encrypt Master key, Database Master Key and Certificate.

*2) Validation*

Validation describes the ability to provide assurance that a senders identity is true and that a Column, Tablespace or file has not been modified. Encryption can be used to provide validation by making a digital Certificate of the information contained within a database. Upon validation, the user can be reasonably sure that the data came from a trusted person and that the contents of the data have not been modified.

*3) Data protection*

Probably the most widely-used application of transparent encryption is in the area of data protection. The information that a business owns is invaluable to its productive operation; consequently, the protection of this information is very important. For people working in small offices and home offices, the most practical uses of transparent encryption for data protection are column, Tablespace and files encryption. This information protection is vital in the event of theft of the computer itself or if an attacker successfully breaks into the system. The encryption and decryption process is performed at the SQL layer, completely transparent to applications and users.

*C. Limitation of Transparent data encryption*

- Transparent data Protection does not provide encryption across communication channels.
- When enabling Transparent data Protection, you should immediately back up the certificate and the private key associated with the certificate. If the certificate ever becomes unavailable or if you must restore or attach the database on another server, you must have backups of both the certificate and the private key or you will not be able to open the database.
- The encrypting certificate or Asymmetric should be retained even if Transparent data Protection is no longer enabled on the database. Even though the database is not encrypted, the database encryption key may be retained in the database and may need to be accessed for some operations.
- Altering the certificates to be password-protected after they are used by Transparent data Protection will cause the database to become inaccessible after a restart.

IV. TRANSPARENT DATA ENCRYPTION IN MICROSOFT SQL SERVER 2008

In Microsoft SQL Server 2008 Transparent Data Encryption of the database file is performed at the page level. The pages in an encrypted database are encrypted before they are written to disk and decrypted when read into memory. Transparent data Protection does not increase the size of the encrypted database.

*A. Architecture of Transparent Data Encryption*

In Microsoft SQL Server 2008 Transparent Data Encryption first upon we need to Create a master key, then obtain a certificate protected by the master key after that Create a database encryption key and protect it by the certificate and Set the database to use encryption. The following Figure 2 shows the steps of Transparent Data Encryption

In Microsoft SQL Server 2008 Transparent Data Encryption of the database file is performed at the page level. The pages in an encrypted database are encrypted before they are written to disk and decrypted when read into memory. Service Master key is created at a time of SQL Server setup, DPAPI encrypts the Service Master key. Service Master key encrypts Database Master key for the Master Database. The Database Master key of the master Database Creates the Certificate then the certificate encrypts the database encryption key in the user database. The entire database is secured by the Database Master key of the user Database by using Transparent Database encryption.

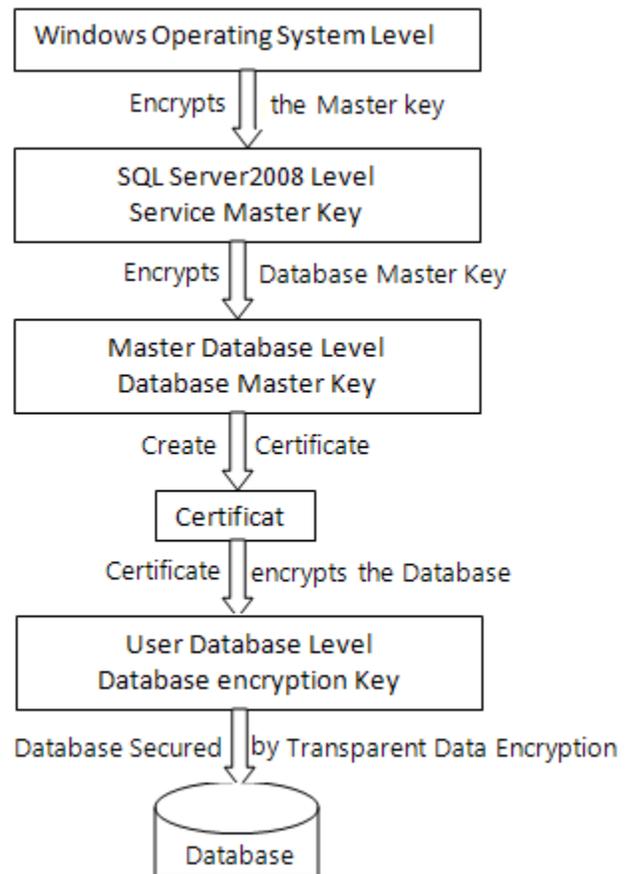

Figure 2 Microsoft SQL Server 2008 Transparent Data Encryption

V. CREATE DATABASE USING MICROSOFT SQL SERVER 2008

Creating a database that specifies the data and transaction log files, the following code 1 to create database name "Sales"

*A. Code 1*

```
USE master;
GO
    CREATE DATABASE Sales
    ON
    ( NAME = Sales_dat,
    FILENAME = 'C:\Program Files\Microsoft SQL Server\MSSQL10_50.MSSQLSERVER\MSSQL\DATA\saledat.mdf',
    SIZE = 10,
    MAXSIZE = 50,
    FILEGROWTH = 5 )
    LOG ON
    ( NAME = Sales_log,
    FILENAME = 'C:\Program Files\Microsoft SQL Server\MSSQL10_50.MSSQLSERVER\MSSQL\DATA\salelog.ldf',
    SIZE = 5MB,
    MAXSIZE = 25MB,
    FILEGROWTH = 5MB ) ;
GO
```

*B. Description of above Code 1*

CREAT DATABASE command is used to create database in SQL Server 2008, Sales is a name of Database. In these example we have created one data file name "Sales_dat' and one lof file named "Sales_log" with specified size

## VI. CREATION OF ENCRYPTION AND DECRYPTION OF DATABASE

The following code 2 illustrates encrypting and decrypting the Sales database using a certificate installed on the server named MySalesCert.

*A. Code 2*

```
USE master;
GO
        CREATE MASTER KEY ENCRYPTION BY
PASSWORD = '<writeanypasswordhere>';
    go
            CREATE CERTIFICATE
        MySalesCert WITH SUBJECT = 'It is my
        Certificate';
    go
    USE Sales;
    GO
            CREATE DATABASE ENCRYPTION
KEY
            WITH ALGORITHM = AES_128
            ENCRYPTION BY SERVER
CERTIFICATE MySalesCert;
        GO
        ALTER DATABASE Sales
        SET ENCRYPTION ON;
GO
```

*B. Description of Code 2*

The Command CREATE MASTER KEY ENCRYPTION BY PASSWORD is used to create Master key encryption with password here user can assign any password. CREATE CERTIFICATE is used to create certificate as MySalesCert, WITH SUBJECT is used to assign any subject as "It is my Certificate" for the database created in code 1 "Sales". The Command CREATE DATABASE ENCRYPTION KEY WITH ALGORITHM = AES_128 it is a encryption key. By using ENCRYPTION BY SERVER CERTIFICATE command assigning the certificate "MySalesCert" to the Server. By using the command SET ENCRYPTION keeping it ON.

## VII. CONCLUSIONS

Transparent Data Encryption plays an especially important role in safeguarding data in transit. Microsoft SQL Server 2008 Transparent Data Encryption protects sensitive data on disk drives and backup media from unauthorized access, helping reduce the impact of lost or stolen media. The Transparent Data Encryption are developed under the present work has been successfully, created, implemented and thoroughly tested. We used this Transparent Data Encryption on few computer we found it to be very effective. The Transparent Data Encryption based on the study of various security measures were planned, designed and developed using techniques and tools described earlier. Transparent Data Encryption provides a highly configurable environment for application development. User can use Transparent Data Encryption to create both single-machine and networked environments in which developers can safely try out. Such a setup do not requires additional infrastructure and easily caters to the needs of beginners as well as advanced learners on one premise.

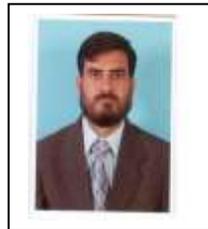

AUTHOR PROFILE

**Dr. Anwar Pasha Deshmukh:** Working as a Lecturer, College of Computer and Information Technology, University of Tabuk, Tabuk, Saudi Arabia. PhD (Computer Science) July 2010, MPhil (Information Technology) 2008, M Sc (Information Technolgy) 2003 from Dr. Babasaheb Ambedkar Marathwada University, Aurangabad. India.

**Dr Riyazuddin Qureshi:** Working as a Assistant Professor, Community College, Jazan University,

Jazan, Kingdom of Saudi Arabia. PhD(Management Science) February 2008, M. Com May 1993, MCA May 1998, Dr. Babasaheb Ambedkar Marathwada University, Aurangabad.